\documentclass{llncs}
\usepackage{llncsdoc}
\usepackage{times}
\usepackage{epstopdf}

\usepackage{graphicx}
\usepackage{subfigure}

\title{GeoMFree$^{3D}$: An Under-Development Meshfree Software Package for 
	Geomechanics}

\author{Gang Mei\inst{1} \and Nengxiong Xu\inst{1} \and Liangliang Xu\inst{1} \and Yazhe Li\inst{1}}

\institute{
	School of Engineering and Technology,\\
	China University of Geosciences (Beijing), 100083, Beijing, China\\
	\email{\{gang.mei; xunengxiong; liang.xu; liyazhe\}@cugb.edu.cn}
}

\begin{document}

\maketitle
\begin{abstract}
This paper briefly reports the GeoMFree$^{3D}$, a meshfree / meshless 
software package designed for analyzing the problems of large deformations 
and crack propagations of rock and soil masses in geotechnics. The 
GeoMFree$^{3D}$ is developed based on the meshfree RPIM, and accelerated by 
exploiting the parallel computing on multi-core CPU and many-core GPU. The 
GeoMFree$^{3D}$ is currently being under intensive developments. To 
demonstrate the correctness and effectiveness of the GeoMFree$^{3D}$, 
several simple verification examples are presented in this paper. Moreover, 
future work on the development of the GeoMFree$^{3D}$ is introduced. 
\end{abstract}


\section{Introduction}

In geotechnics, the large deformations of rock and soil masses commonly 
occur in various geo-disasters such as landslides, debris flows, rock 
collapses, and ground subsidence. Moreover, when analyzing the stability of 
rock or soil slopes, the distribution and propagation of natural cracks is 
one of the most crucial issues that needs to be considered. To understand 
the mechanisms behind the above-mentioned geo-disasters, physical 
experiments and numerical investigations are commonly employed in practice. 

The large-deformations and crack propagations of rock and soil masses have 
been examined using various mesh-based or meshfree numerical methods \cite{01,02,03,04,05,06}. 
When employing those mesh-dependent numerical analysis 
methods such as Finite Element Method (FEM), Finite Volume Method (FVM), 
Finite Difference Method (FDM) to analyze the large deformation or crack 
propagation in geotechnics, the mesh element would generally be distorted or 
need to be broken. Mainly motivated by addressing those problems mentioned 
above occurring in mesh-based methods, the meshfree / meshless methods such 
as SPH, MLPG, LBIM, EFG, RPIM are proposed; see several excellent reviews \cite{07,08,09}. 

Recently, there are several meshfree software packages that have been 
developed or are being under development. For example, Hsieh and Pan 
described the Essential Software Framework for Meshfree methods (ESFM) \cite{10}. Cercos-Pita \cite{11} introduced the AQUAgpusph, a 
new free 3D SPH solver accelerated with OpenCL. Sinaie et \textit{al} \cite{12} 
presents the implementation of the material point method (MPM) using Julia. 
Winkler et \textit{al} \cite{13} introduced gpuSPHASE, a shared memory caching 
implementation for 2D SPH using CUDA. Vanaverbeke et \textit{al} \cite{14} 
presented GRADSPMHD, a completely Lagrangian parallel magnetohydrodynamics 
code based on the SPH formalism. Zhang et \textit{al} \cite{15,16,17} developed the 3D 
explicit parallel MPM code, MPM3D.

This short paper briefly reports the GeoMFree$^{3D}$, a meshfree / meshless 
software package for Geomechanics. The objective for developing the 
GeoMFree$^{3D}$ is to numerically analyze large deformations \cite{18,19,20},
and crack propagations \cite{21,22} of rock and soil masses in 
geomechanics. The package GeoMFree$^{3D}$ is currently under intensive 
developments. The underlying algorithm behind the GeoMFree3D is the Radial 
Point Interpolation Method (RPIM) proposed by Liu G.R. \cite{23,24}, . In 
addition, to improve the computational efficiency when analyzing large-scale 
problems \cite{25}, the GeoMFree3D is parallelized on multi-core CPU 
and many-core GPU using the OpenMP \cite{26} and CUDA  \cite{27}, 
respectively.

\section{GeoMFree$^{3D}$}

The GeoMFree$^{3D}$ is a meshfree software package designed for numerically 
analyzing the large deformations and crack propagations of rock and soil 
masses in geomechanics. The GeoMFree$^{3D}$ is currently capable of 
analyzing linear and nonlinear static problems, and is being developed for 
addressing dynamic problems. The GeoMFree$^{3D}$ is written in C/C++ and 
accelerated by exploiting the parallel computing on multi-core CPU and 
many-core GPU.

\begin{figure}[!h]
	\centering
	\includegraphics[width=0.78\linewidth]{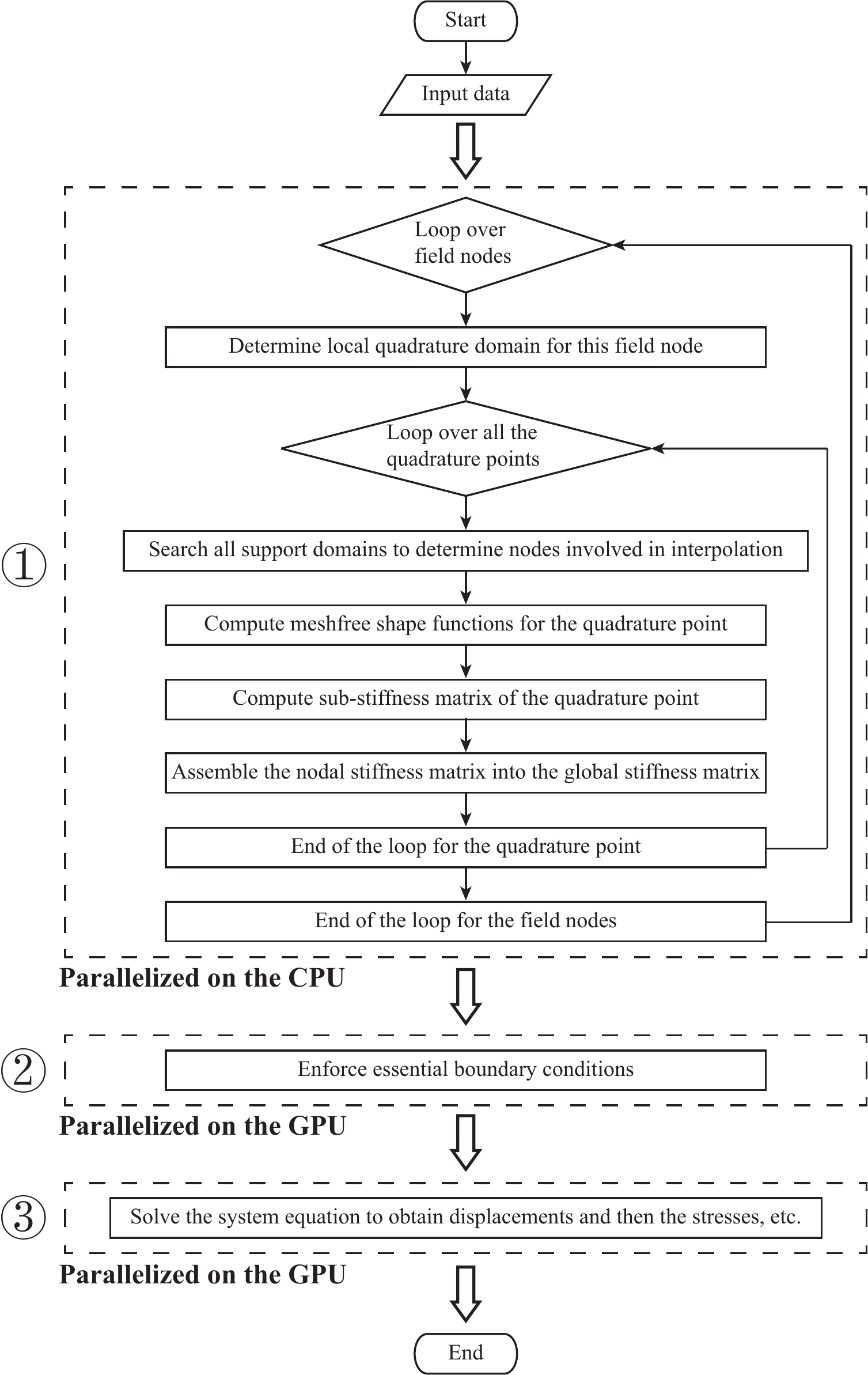}
	\caption{Process of our meshfree software package GeoMFree$^{3D}$}
	\label{fig1}
\end{figure}

The process of numerical modeling using the GeoMFree$^{3D}$ is illustrated 
in Figure \ref{fig1}. There are three major stages in the GeoMFree$^{3D}$. The first 
stage is to assemble the global stiffness matrix by looping over all field 
nodes to create the element stiffness matrix of each field node. The second 
is to enforce the boundary conditions. And the third is to solve the system 
equation to obtain displacements and then the stresses, etc.

To improve the computational efficiency, the first stage of assembling the 
global stiffness matrix is parallelized on multi-core CPU using the API 
OpenMP \cite{26}. The meshfree RPIM is inherently suitable to be 
parallelized since there is no data dependencies between the forming of any 
two element stiffness matrices for any pair of field nodes. That is, the 
assembly of the element stiffness matrix of one field node is completely 
independent of that for another field node. Therefore, we can allocate $n$ 
threads on the multi-core CPU; and each thread is responsible for assembling 
the element stiffness matrix for one field node. In this case, the assembly 
of element stiffness matrices for $n$ field nodes can be conducted 
concurrently. This is the essential idea behind parallelizing the assembly 
of global / system stiffness matrix on multi-core CPU. 

Similarly, to enhance the computational efficiency, the second stage of 
enforcing the boundary condition can also be parallelized. More 
specifically, we adopt the penalty function method to enforce the 
displacement boundary conditions. This procedure is performed in parallel on 
the many-core GPU. Assuming there are $m$ field nodes on the displacement 
boundary, and we can allocate $m$ GPU threads to enforce the displacement 
boundary conditions for the $m$ field nodes concurrently, where each thread 
takes responsibility for enforcing the displacement boundary condition for 
one boundary field node. 

The final stage is the solving of system equations to obtain nodal 
displacements and then the stresses. In meshfree RPIM, the assembled global 
stiffness matrix is large, sparse, and asymmetric. When analyzing 
large-scale problems and requiring a large number of field nodes, the global 
system matrix could be very large. To improve the computational efficiency 
in solving system equations, we have employed the \texttt{cuSparse} and \texttt{cuSolver} 
library integrated in CUDA \cite{27} to solve the  system of 
equations. 

 \section{Verification}

This section will present several computational examples to verify the 
validation and features of the reported meshfree software package 
GeoMFree$^{3D}$.

\subsection{Example 1: Stresses of a Cubic Domain}

\begin{figure}[htbp]
	\centering
	\subfigure[]{
		\label{fig2a}       
		\includegraphics[width=0.45\linewidth]{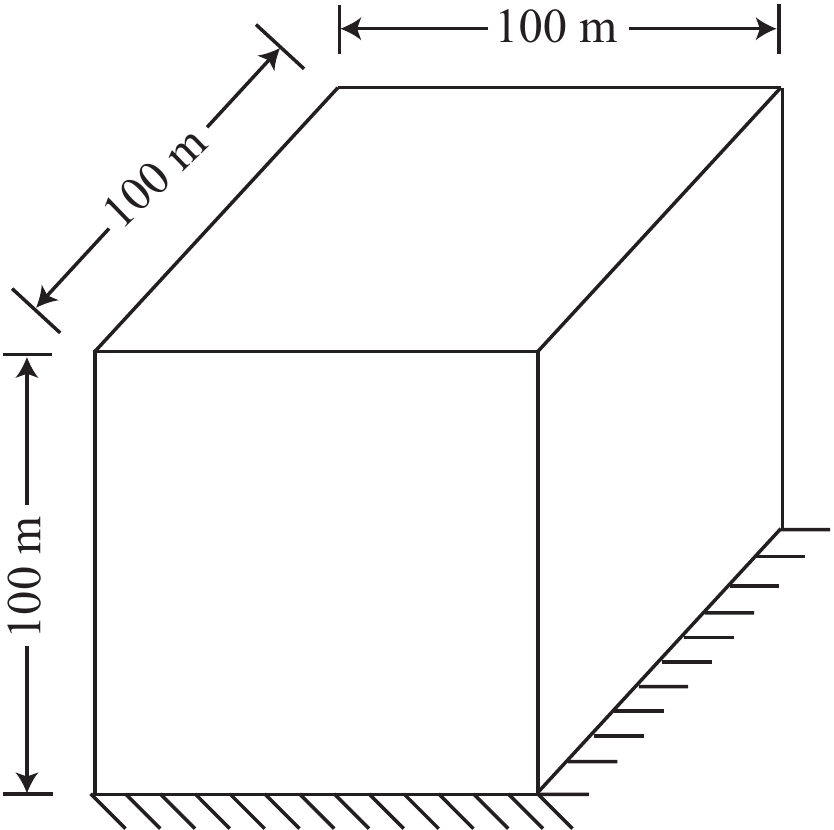}
	}
	\subfigure[]{
		\label{fig2b}       
		\includegraphics[width=0.45\linewidth]{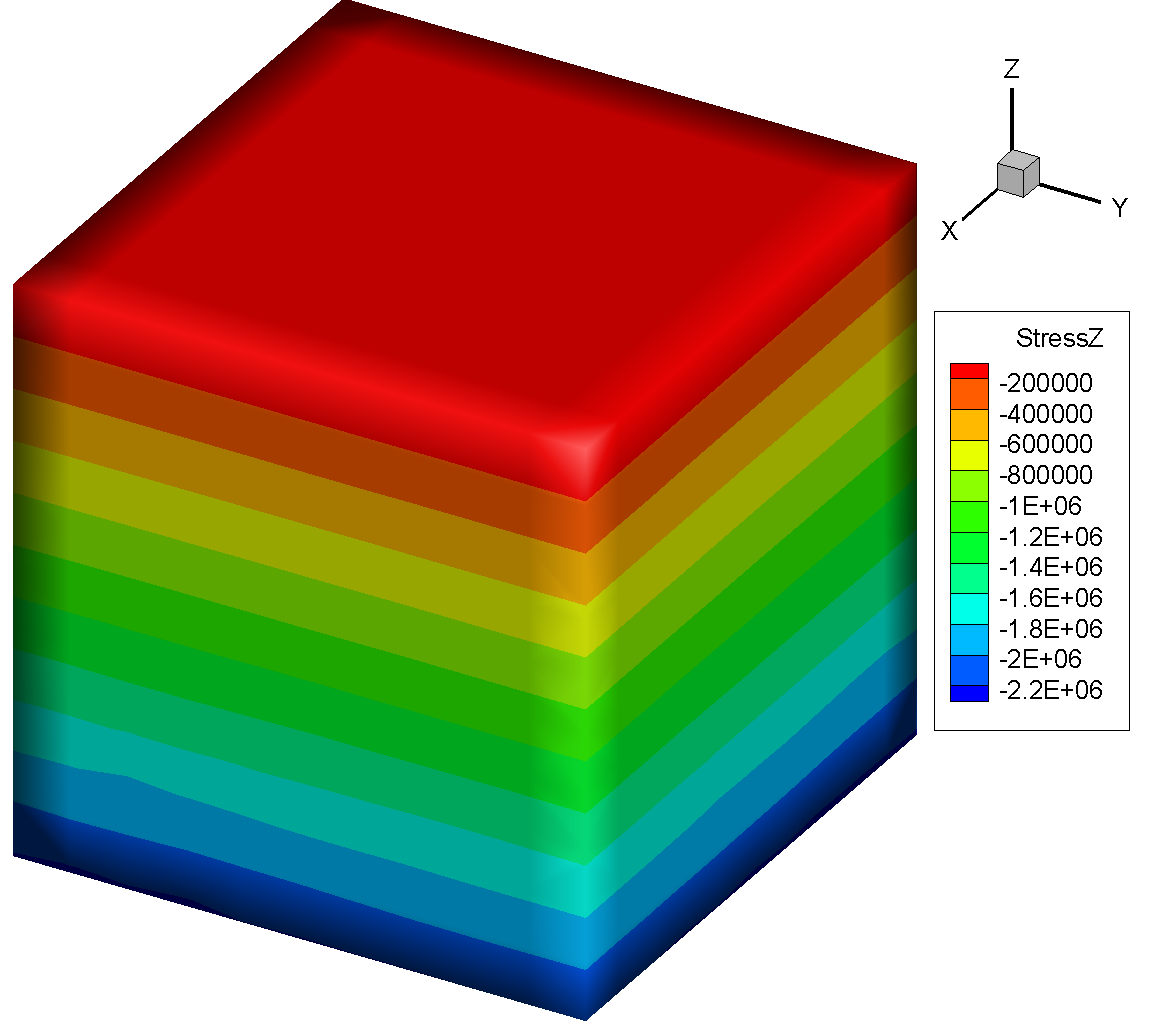}
	}
	\caption{Verification example 1: stresses of a cubic domain. (a) 
		Computational model of a cubic domain; (b) Stresses calculated by using our 
		package GeoMFree$^{3D}$.}
	\label{fig2}       
\end{figure}

First, to verify the correctness of the GeoMFree$^{3D}$, we specifically 
calculate the distribution of displacements and stresses of a cubic domain; 
see Figure \ref{fig2}. In this quite simple verification example, only the force of 
gravity is considered and there are no other forces. The density of the cube 
is set as 2600 kg / m$^{3}$. The stress on the bottom of the cube can be 
theoretically calculated, and is noted as the \textit{theoretical} result. In contrast, we can 
also numerically calculate the nodal stress on the bottom using our meshfree 
package GeoMFree$^{3D}$ which is noted as the \textit{numerical} result. Then, by comparing 
the theoretical result to the numerical counterpart, we could validate the 
correctness of the GeoMFree$^{3D}$. 

The theoretical nodal stress on the bottom of the cube is 2.548 MPa, while 
the numerically calculated one is 2.376 MPa. There is a slight difference 
between the theoretical and numerical results. And thus, we can conclude 
that the correctness of the GeoMFree$^{3D}$ has been verified, although the 
employed verification example is extremely simple. 

\subsection{Example 2: Displacements of a Cantilever Beam with Crack}

\begin{figure}[htpb]
	\centering
	\subfigure[Computational model of a Cantilever beam with a crack]{
		\label{fig3a}       
		\includegraphics[width=0.7\linewidth]{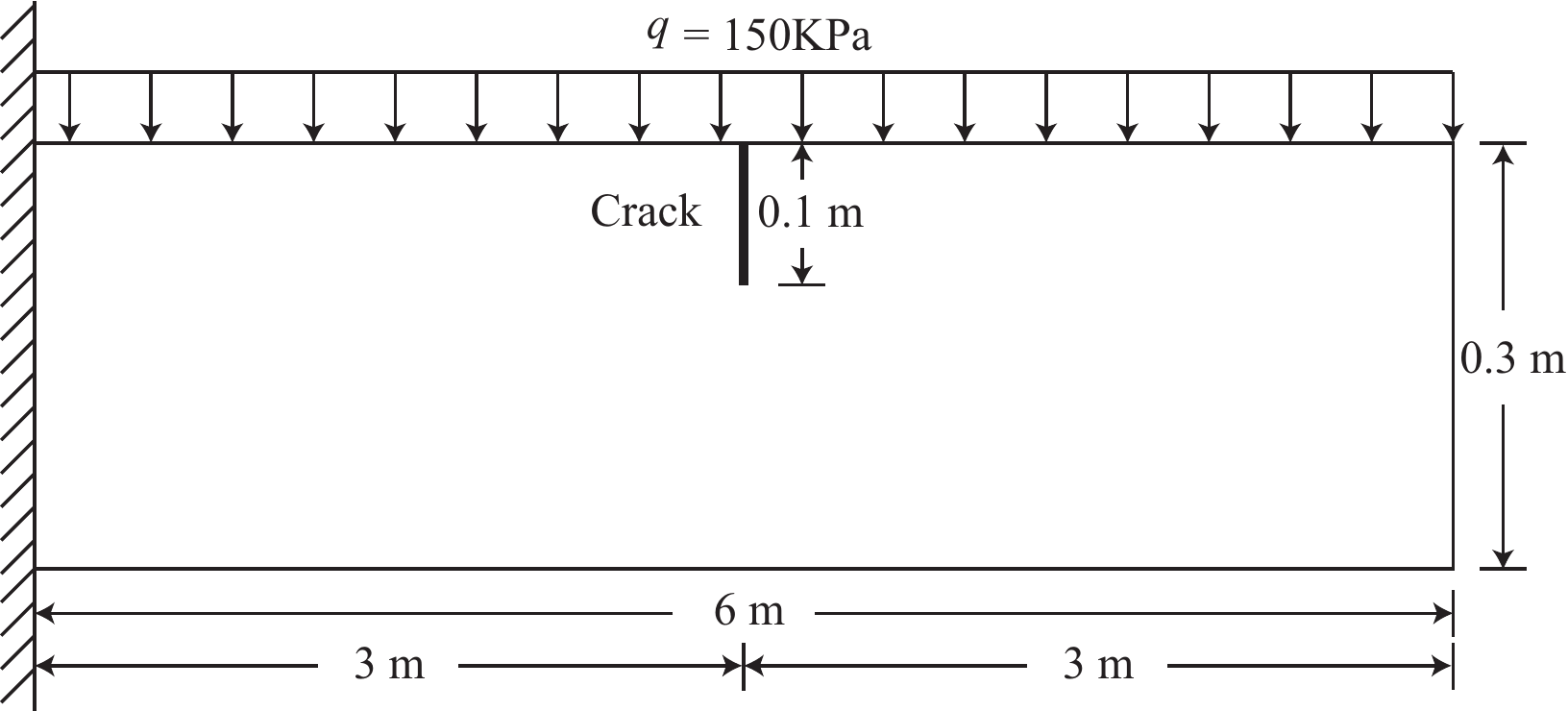}
	}
	\subfigure[Displacements calculated by using the software FLAC$^{3D}$]{
		\label{fig3b}       
		\includegraphics[width=0.75\linewidth]{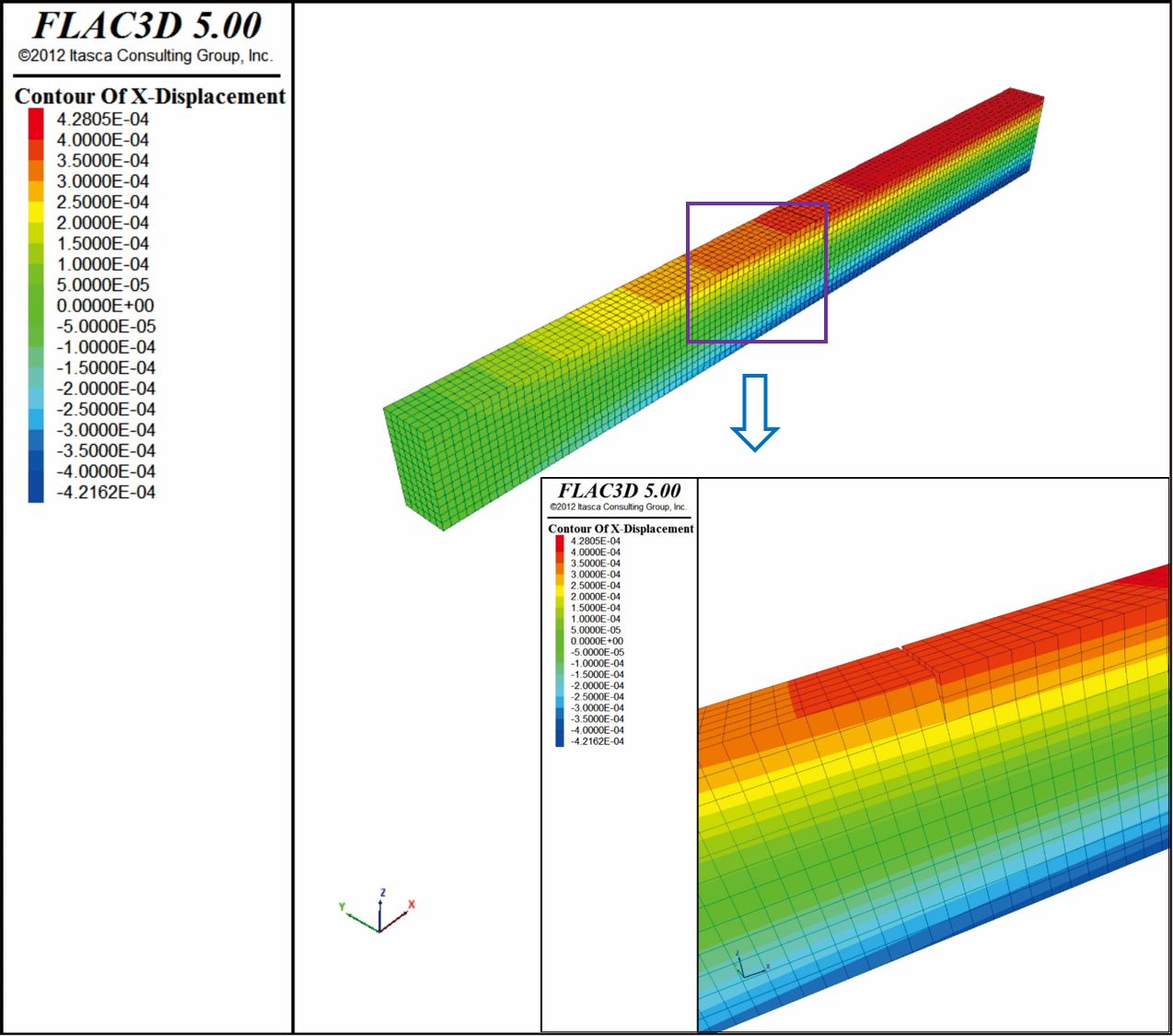}
	}
	\subfigure[Displacements calculated by using our package GeoMFree$^{3D}$]{
		\label{fig3c}       
		\includegraphics[width=0.75\linewidth]{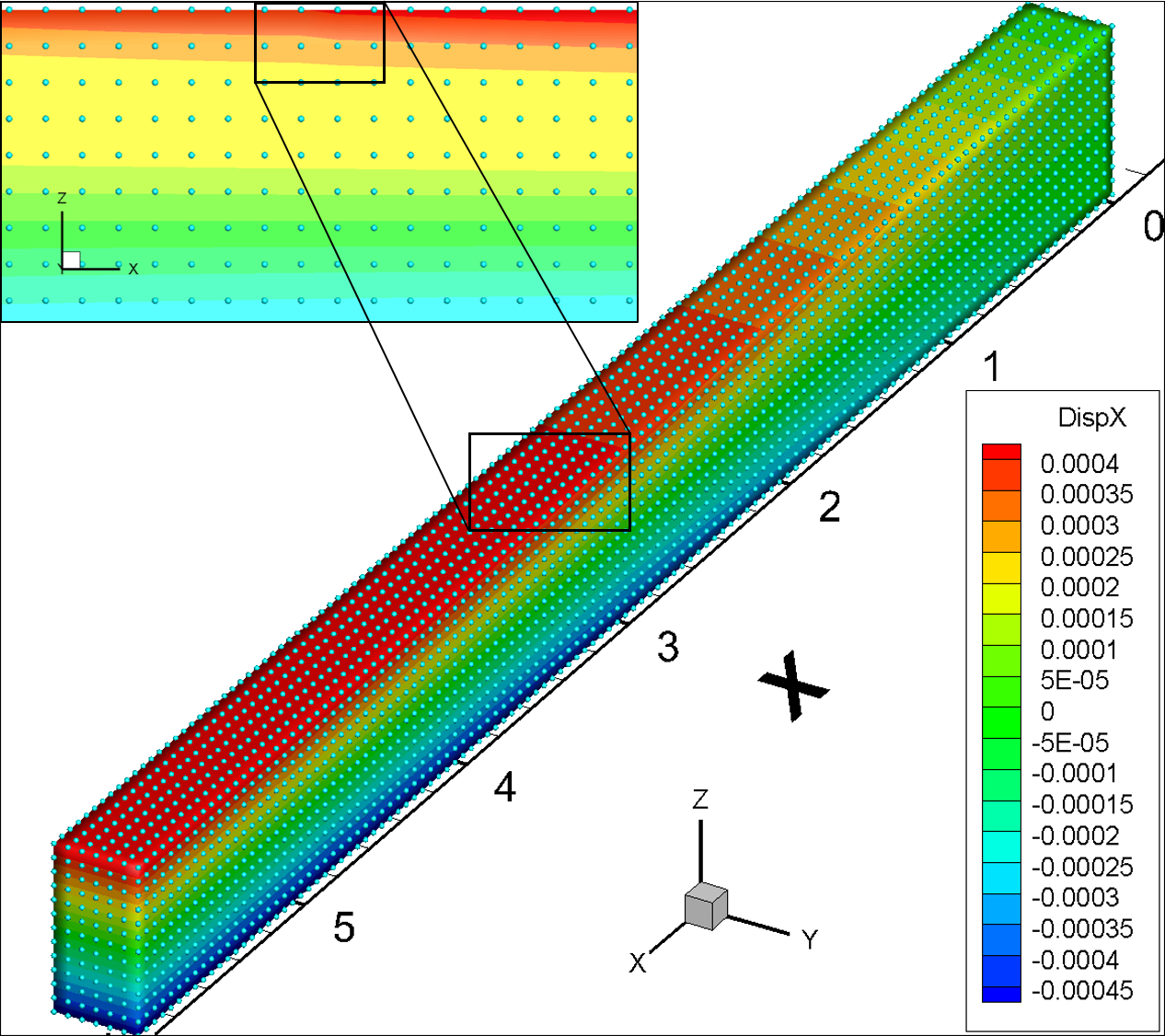}
	}
	\caption{Verification example 2: displacements of a Cantilever beam with crack}
	\label{fig3}
\end{figure}

To further verify the effectiveness of our GeoMFree$^{3D}$, we have employed 
it to calculate the displacement and stress field of a Cantilever beam with 
a crack; see Figure \ref{fig3}. Moreover, we have also computed the displacements and 
stresses of the beam using a FDM numerical software FLAC$^{3D}$; see Figure 
\ref{fig3b}. The numerical results calculated by our package GeoMFree$^{3D}$ and 
the commercial numerical software FLAC$^{3D}$ are almost the same; see 
Figure \ref{fig3b} and \ref{fig3c}. And this indicates that currently our package 
GeoMFree$^{3D}$ is capable of analyzing the very simple cases of crack 
propagation. In the near future, we hope that the GeoMFree$^{3D}$ can be 
used to model and simulate dynamic crack propagations in three-dimensions. 

\subsection{Example 3: Displacements of a Simplified Slope}

\begin{figure}[htbp]
	\centering
	\subfigure[Geometrical model of a simplified slope ]{
		\label{fig4a}       
		\includegraphics[width=0.7\linewidth]{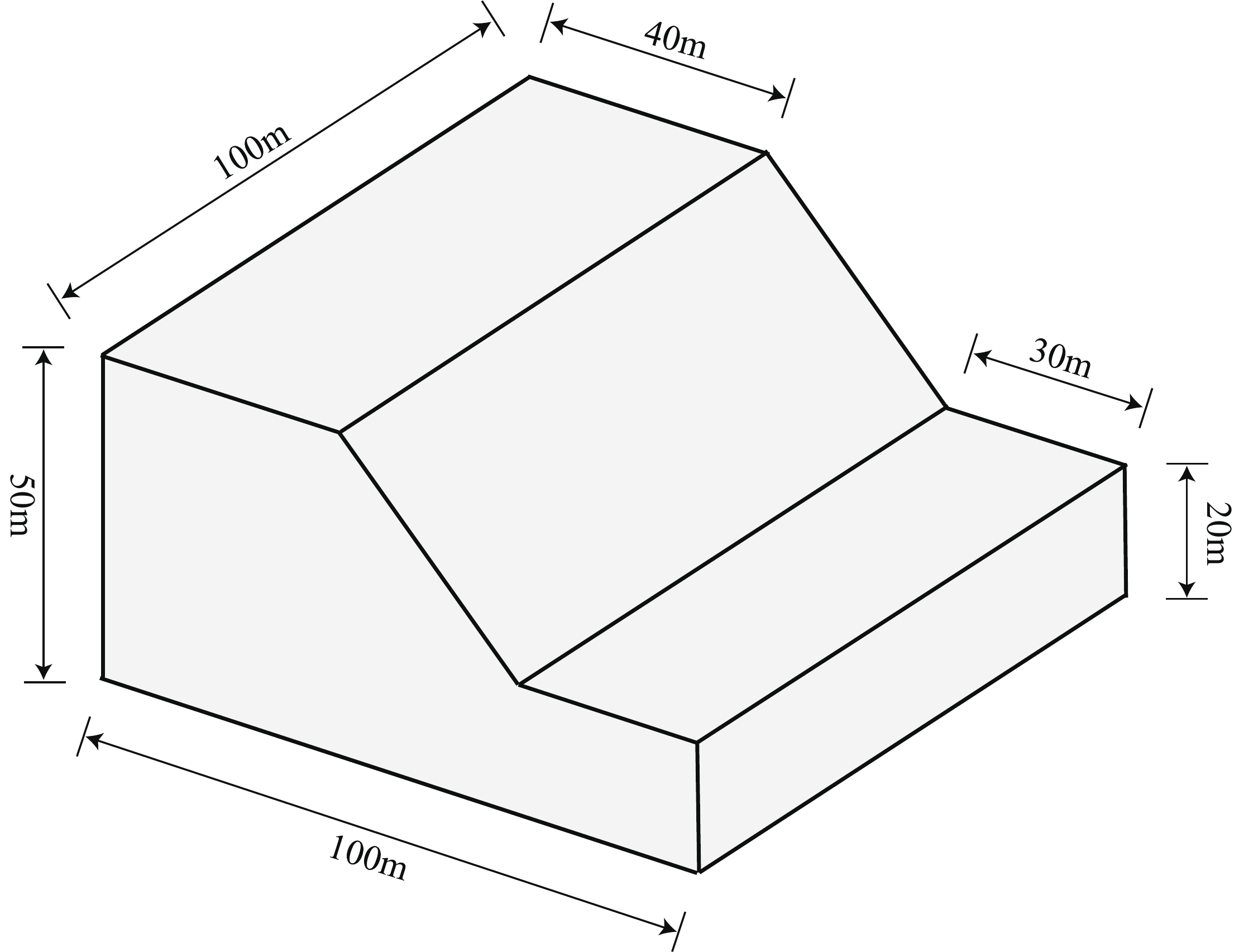}
	}
	\subfigure[Displacements calculated by using the software FLAC$^{3D}$]{
		\label{fig4b}       
		\includegraphics[width=0.75\linewidth]{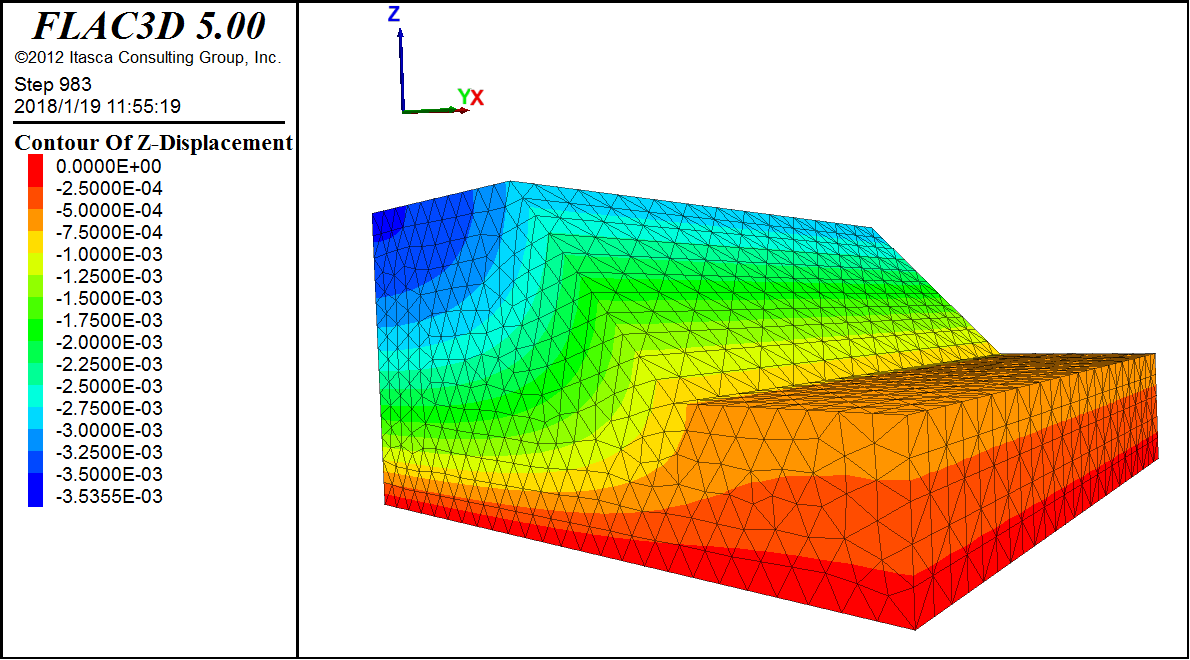}
	}
	\subfigure[Displacements calculated by using our package GeoMFree$^{3D}$]{
		\label{fig4c}       
		\includegraphics[width=0.75\linewidth]{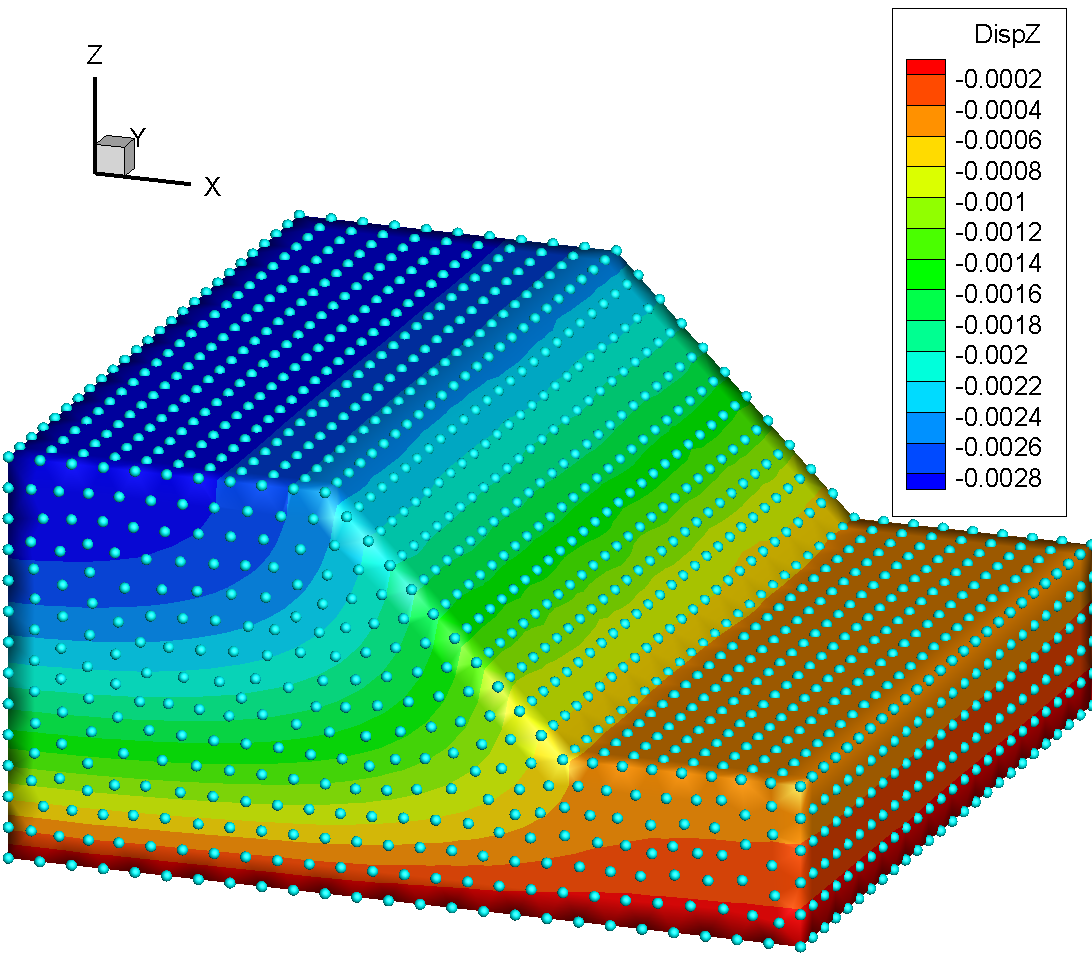}
	}
	\caption{Displacements of a simplified slope when employing 
		regularly-distributed field nodes}
	\label{fig4}
\end{figure}

\begin{figure}[!h]
	\centering
	\subfigure[Displacements calculated by using the software FLAC$^{3D}$]{
		\label{fig5a}       
		\includegraphics[width=0.75\linewidth]{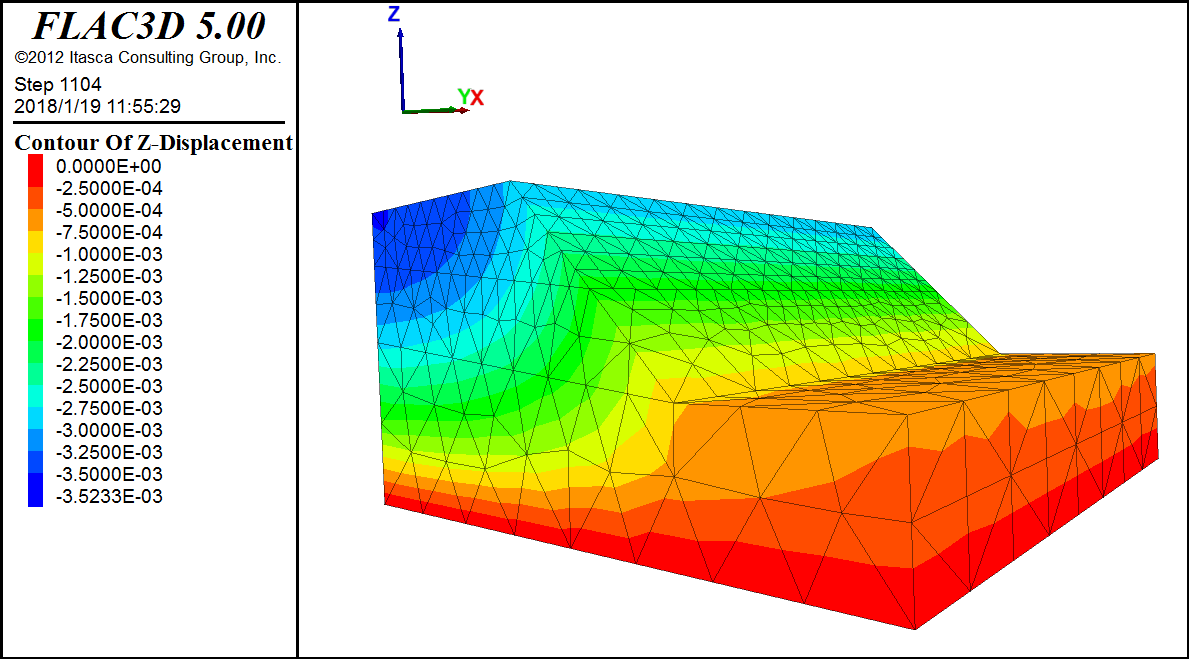}
	}
	\subfigure[Displacements calculated by using our package GeoMFree$^{3D}$]{
		\label{fig5b}       
		\includegraphics[width=0.75\linewidth]{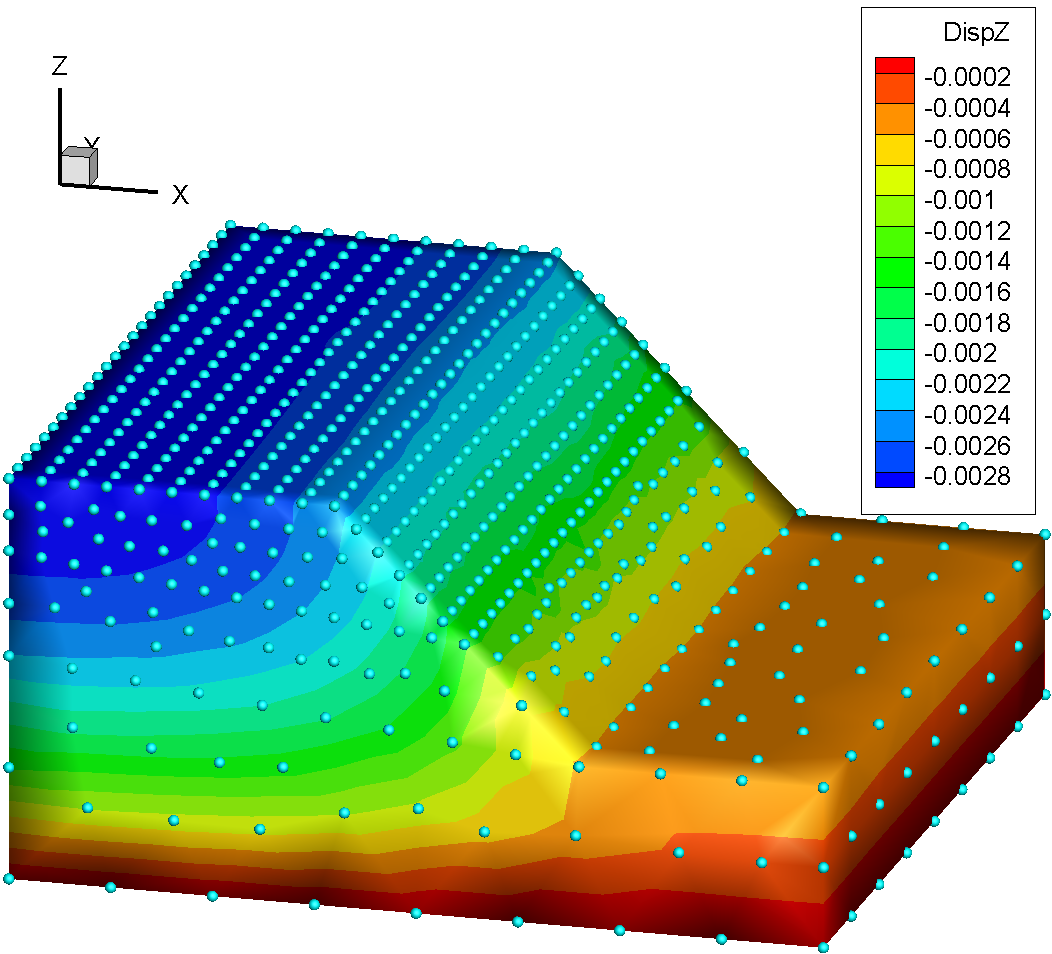}
	}
	\caption{Displacements of a simplified slope when employing 
		irregularly-distributed field nodes}
	\label{fig5}
\end{figure}

As having been introduced several times, the motivation why we are 
developing our meshfree package GeoMFree$^{3D}$ is that: we hope to employ 
one of the meshfree methods, i.e., the RPIM, to well model and simulate the 
large deformations and crack propagations of rock and soil masses. 
Currently, the GeoMFree$^{3D}$ cannot be used to model the continuously 
developed large deformations of rock or soil masses. But it can be used to 
calculate the displacements of a simplified slope; see Figure \ref{fig4} and Figure 
\ref{fig5}. And we are working on analyzing the stability of slopes using the 
GeoMFree$^{3D}$ based upon the Strength Reduction Method (SRM). 

In meshfree methods, the study domain is discretized with a set of field 
nodes. These field nodes could be (1) \textit{regularly} or (2) \textit{irregularly} distributed in the domain. 
The patterns of nodal distributions are of strong influence on both the 
computational accuracy and efficiency. To verify the flexibility of the 
GeoMFree3D for addressing problems with regular or irregular discretization, 
we have decomposed a simplified slope model with: (1) regular nodes (Figure 
\ref{fig4b}) and (2) irregular nodes (Figure \ref{fig5a}). We then calculated the 
displacements of the above two models using our package GeoMFree$^{3D}$, and 
also compared the results calculated by the GeoMFree$^{3D}$ to those by the 
commercial software FLAC$^{3D}$. 

The numerical results illustrated in Figure 4 and Figure 5 indicate that: 
(1) our package GeoMFree$^{3D}$ is capable of analyzing the problems with 
regular or irregular nodal distributions; (2) our package GeoMFree$^{3D}$ 
can be used to address the problems with relatively complex geometric 
domains and boundaries.

\section{Conclusion and Future Work}

A meshfree software package, GeoMFree$^{3D}$, has been briefly introduced in 
this paper. The package GeoMFree$^{3D}$ is designed for the numerical 
investigation of large-deformations and crack propagations of rock and soil 
masses in geotechnics. The GeoMFree$^{3D}$ is developed based on the RPIM, 
and is currently under intensive developments. To validate the effectiveness 
of the introduced GeoMFree$^{3D}$, several verifications have been 
conducted. The verification examples have demonstrated that the current 
version of GeoMFree$^{3D}$ is capable of analyzing the deformation of simple 
study domains. 

The GeoMFree$^{3D}$ is currently under intensive developments. We are 
focusing on improving the computational efficiency by developing accurate 
and efficiency meshfree shape functions \cite{28,29,30}, for example, the 
parallel RBF \cite{31}, MLS \cite{32}, and Shepard \cite{33,34}
interpolations. Currently, we are also aiming at numerically modeling the 
crack propagation of multiple tensile and shear cracks of rock masses. In 
future, we hope that: the GeoMFree$^{3D}$ can be used to (1) model the 
large-deformations of strata induced by underground mining and (2) analyze 
the stability of jointed rock slopes via modeling the very complex crack 
propagations of rock masses.

\section*{Acknowledgements} This research was supported by the Natural Science Foundation of China 
(Grant Numbers 41772326 and 11602235), and the Fundamental Research Funds 
for the Central Universities. The authors would like to thank the editor and 
reviewers for their contributions on the paper.

\bibliographystyle{splncs}%

\bibliography{MyRef}

\end{document}